\documentclass[aps,pre,reprint,superscriptaddress,longbibliography]{revtex4-1}

\usepackage{amssymb}
\usepackage{amsmath}
\usepackage{cases}
\usepackage{subfigure}
\usepackage{graphicx,color}
\definecolor{r}{rgb}{1,0,0}   
\definecolor{g}{rgb}{0,1,0}   
\definecolor{b}{rgb}{0,0,1}

\begin{document}


\title{A simply solvable model capturing the approach to statistical self-similarity for the diffusive coarsening of bubbles, droplets, and grains}



\author{Anthony T. Chieco}
\affiliation{Department of Physics \& Astronomy, University of Pennsylvania, Philadelphia, PA 19104-6396, USA}

\author{Douglas J. Durian}
\affiliation{Department of Physics \& Astronomy, University of Pennsylvania, Philadelphia, PA 19104-6396, USA}
\affiliation{Center for Computational Biology, Flatiron Institute, Simons Foundation, New York, NY 10010, USA}



\date{\today}

\begin{abstract}
Aqueous foams and a wide range of related systems are believed to coarsen by diffusion between neighboring domains into a statistically self-similar scaling state, after the decay of initial transients, such that dimensionless domain size and shape distributions become time independent and the average grows as a power law. Partial integrodifferential equations for the time evolution of the size distribution for such phase separating systems can be formulated for arbitrary initial conditions, but these are cumbersome for analyzing data on non-scaling state preparations.  Here we show that essential features of the approach to the scaling state are captured by an exactly-solvable ordinary differential equation for the evolution of the average bubble size.  The key ingredient is to characterize the bubble size distribution approximately, using the average size of all bubbles and the average size of the critical bubbles, which instantaneously neither grow nor shrink.  The difference between these two averages serves as a proxy for the width of the size distribution. To test our model, we compare with data for quasi-two dimensional dry foams created with three different initial amounts of polydispersity. This allows us to readily identify the critical radius from the average area of six-sided bubbles, whose growth rate is zero by the von~Neumann law. The growth of the average and critical radii agree quite well with exact solution, though the most monodisperse sample crosses over to the scaling state faster than expected. A simpler approximate solution of our model performs equally well. Our approach is applicable to 3d foams, which we demonstrate by re-analyzing prior data, as well as to froths of dilute droplets and to phase separation kinetics for more general systems such as emulsions, binary mixtures, and alloys.
\end{abstract}


\maketitle




\section{Introduction}

Two-phase systems consisting of compact domains, whether dense or dilute, tend to evolve with time due to the surface tension energy of the interface between the discrete domains and the continuous interstitial medium. Even in cases where direct domain-domain coalescence is forbidden, domain sizes can still evolve by diffusion of their material across the interstitial medium. With time, smaller domains shrink while larger domains grow and this gives a continual lowering of the total interfacial area - as befits a process driven by surface tension. This is variously called ``Ostwald ripening" or ``diffusive coarsening" and occurs whether either of the phases is liquid, solid, amorphous, crystalline, macroscopic, nanometric, two-dimensional, or three-dimensional.  Historically, the pioneering theoretical works by Mullins \cite{Mullins56}, von~Neumann \cite{VonNeumann}, Lifshitz \& Slyozov \cite{Lifshitz1961}, and Wagner \cite{Wagner1961} were presented in the context of grain growth for metallic alloys. But they apply equally well to bubbles or droplets in a liquid where the microstructure is well understood \cite{Lemlich1978, Markworth1985, Stavans93, Taylor1998}. Furthermore, the case of dry foams, where the bubbles essentially fill space, is ideal for general study of collective effects in densely-packed systems because the single-bubbled growth laws are known exactly in two \cite{Mullins56, VonNeumann} and three dimensions \cite{MacPhersonSrolovitz2007} in terms of the geometry of individual bubbles.

While the pioneering papers and reviews cited above focus primarily on individual-domain growth laws and behavior in the statistically self-similar scaling state, usually in the context of specific systems, our focus here is on the decay of transient memory of the initial preparation as the system approaches scaling. Prior studies are primarily by experiment or simulation  \cite{Venzl1983, VoorheesJSP85, Beenakker86, GlazierGrossStavans87, MarderPRA87, BeenakkerPRA88, FeitosaDurian06, GlazierGraner10, DuplatJFM2011, BaroudPRL2019, ZimnyakovCSA2019}, because treatment of the full size distribution requires numerical solution of partial integrodifferential equations. As of yet there is no ready means to analyze the decay of transients on approach to scaling.  Here, we demonstrate how to do so by accounting for the domain size distribution approximately - in terms of only the average and critical domain sizes.

To begin, we review prior models and then lay out the ingredients needed for our approach. After deriving relatively simple ordinary differential \textit{average coarsening equations}, we present perturbative and exact solutions. Then we report on experimental tests of the ingredients of our model and its predictions using quasi-two dimensional dry foams of bubbles squashed between parallel plates. Lastly, bolstered by this success, we use our simple perturbative solutions to re-analyze prior data on three-dimensional foams from Refs.~\cite{DurianWeitzPine91b, GlazierGraner10, ZimnyakovCSA2019}.

\section{Prior Models}

To be concrete, we couch discussions in the context of aqueous foams, which coarsen by the diffusion of gas across the liquid between neighboring bubbles of different pressure \cite{WeaireHutzlerBook, Cantat2013, Langevin2020}.  The $d$-dimensional volume of a gas bubble in an isotropic environment grows or shrinks at rate given by Fick's law as $dR_i^d/dt=D\nabla\varphi R_i^{d-1}$ where $R_i$ is the radius of bubble $i$, $d$ is dimensionality, $D$ is gas diffusivity, $\nabla\varphi$ is the concentration gradient of dissolved gas in the continuous liquid phase at the bubble's surface, $R_i^{d-1}$ is the bubble's surface area, and numerical prefactors are suppressed.  The concentration gradient is set by $\nabla\varphi = H\Delta P/L$ where $H$ is Henry's constant, $\Delta P$ is a pressure difference reflecting the width of the size distribution, and $L$ is a length scale.  The former may be taken from Laplace's law as $\Delta P = \sigma(1/R_c - 1/R_i)$ where $\sigma$ is surface tension and  $R_c$ is the critical radius of bubbles that neither grow nor shrink.  The characteristic length scale for the concentration gradient is $L=R_i$ for dilute droplets in a very wet froth and is $L=\ell$ for compressed bubbles separated by soap films of thickness $\ell$ in a very dry foam. Altogether, in any dimension, the growth laws for individual bubbles are thus of form
\begin{subnumcases}{\label{eq_dRdt1} \frac{dR_i}{dt}=}
	\label{eq_dRdtFroth1}
		\frac{\alpha_w}{R_i}\left( \frac{1}{R_c}-\frac{1}{R_i} \right) & wet froth \\
	\label{eq_dRdtFoam1}
		\alpha\left( \frac{1}{R_c}-\frac{1}{R_i} \right) & dry foam
\end{subnumcases}	
where $\alpha_w$ and $\alpha$ are materials constant for the two cases.  Eq.~(\ref{eq_dRdtFroth1}) is a key ingredient in the Lifshitz-Slyozov/Wagner treatment of grain growth \cite{Lifshitz1961, Wagner1961}. It can be considered exact for the evolution of very dilute spherical droplets of alloy, gas, liquid, etc.  Eq.~(\ref{eq_dRdtFoam1}) is similarly a key ingredient in Lemlich's treatment of coarsening foams \cite{Lemlich1978, Markworth1985}, where the critical radius is $R_c=\langle R^2\rangle/\langle R\rangle$. It is approximate for individual bubbles, in comparison with the exact $d=3$ expression \cite{MacPhersonSrolovitz2007} and the exact $d=2$ von~Neumann law that the rate of area change for a bubble $i$ with $n_i$ sides is $dA_i/dt=K_o(n_i-6)$~\cite{Mullins56, VonNeumann}.  For the latter, note that substituting $A_i \approx \pi R_i^2$ and $n_i \approx 6R_i/R_c$ (Desch's law that perimeter is proportional to side number \cite{Rivier85, ChiuReview}) gives $dR_i/dt \approx (3/\pi)K_o(1/R_c - 1/R_i)$.  Eqs.~(\ref{eq_dRdt1}) have been widely used for disparate systems; \textit{e.g.}~see reviews \cite{VoorheesJSP85, Stavans93, Taylor1998}.  Supplemented with a mass conservation condition, they allow prediction of the evolution of the bubble size distribution. The resulting partial integrodifferential equations are cumbersome and require numerical solution in general.  But, as emphasized by Mullins \cite{Mullins1986}, the distributions tend toward a scaling state of statistical self-similarity such that the shape is time-independent when scaled by the average. For the case of dry 2d foams in the scaling state, the average growth rate can be expressed several ways in terms of different moments of the bubble-size and side-number distributions (see Eqs.~(8,14,15) of Ref.~\cite{ChiecoFSM22}; \textit{e.g.}~the first of these is $d\langle A\rangle/dt = 2K_o[\langle A\rangle^2/\langle A^2\rangle][\langle\langle n\rangle\rangle-6]$ where $\langle\langle n\rangle\rangle \approx 6.8$ is the area-weighted average side number).

In the self-similar scaling state, the critical radius $R_c$ as well as all combinations of moments of the size distribution with units of length must all be proportional to the average bubble radius $\langle R\rangle$. Eqs.~(\ref{eq_dRdt1}) then imply $d\langle R\rangle/dt\propto 1/\langle R\rangle^2$ in the wet limit and $d\langle R\rangle /dt\propto 1/\langle R\rangle$ in the dry limit.  These give the familiar asymptotic power-laws $\langle R(t)\rangle \sim t^{1/3}$ and $\langle R(t)\rangle \sim t^{1/2}$, respectively.  Note that this growth in average radii is accompanied by a decrease in the total number of bubbles and a reduction of total interfacial surface free energy of the system, which drives the process.

\section{Average Coarsening Equations}

While Eqs.~(\ref{eq_dRdt1}) are for individual bubbles, we now suppose a similar form holds for the average bubble radius, $\langle R\rangle$, which for convenience we henceforth denote more simply by $R$.  In particular, we take
\begin{subnumcases}{\label{eq_dRdt} \frac{dR}{dt}=}
		\frac{a_w}{R}\left( \frac{1}{R_c}-\frac{s}{R} \right) & froth \label{eq_dRdtFroth} \\
		a\left( \frac{1}{R_c}-\frac{s}{R} \right) & foam \label{eq_dRdtFoam}
\end{subnumcases}	
where $a_w$ and $a$ are materials constants, slightly different from $\alpha_w$ and $\alpha$, and $s$ is a number that we introduce as a potential fitting parameter because $\langle 1/R\rangle \neq 1/\langle R\rangle$ and also because bubbles also have a distribution of nonspherical shapes. The value of $s$ is expected to be close to one for narrow size and shape distributions. Since $R_c$ is proportional to $R$ in the scaling state, we further suppose that their time derivatives are proportionate at all times:
\begin{equation}
	\frac{dR_c}{dt} = m \frac{dR}{dt}
\label{eq_mdef}
\end{equation}
where $m$ is a dimensionless proportionality constant. While an individual bubble of radius $R_i=R_c$ would neither grow nor shrink, here $R_c$ is the variable for an average property of the distribution for which $dR_c/dt>0$ holds. In effect, we account for a key property of the size distribution by the value of $R_c$ relative to $R$; such a treatment is necessarily approximate, and cannot capture the difference in behavior for distributions that happen to have the same average and critical radii. With these caveats, note that Eqs.~(\ref{eq_dRdt}-\ref{eq_mdef}) imply $d(R_c - mR)/dt = 0$.  As coarsening proceeds, the critical and average radii hence grow in linear relation
\begin{equation}
	R_c(t)=m\left[R(t)-R_o\right] + R_{co}
\label{eq_RcVersusR}
\end{equation}
where $R_o=R(t_o)$ and $R_{co}=R_c(t_o)$ are the average and critical radii at time $t_o$.  Therefore, the assumptions underlying this simplified approach could be tested, and the value of $m$ could be found from experimental or simulation data, by plotting $R_c$ parametrically versus $R$ and fitting to a line.  Furthermore, Eq.~(\ref{eq_RcVersusR}) may be substituted into Eqs.~(\ref{eq_dRdtFroth}) and (\ref{eq_dRdtFoam}), giving
\begin{subnumcases}{\label{eq_dR} \frac{dR}{dt}=}
	\label{eq_dRfroth}
		\frac{a_w}{R}\left[ \frac{1}{ m(R-R_o)+R_{co}}-\frac{s}{R} \right] & froth \\
	\label{eq_dRfoam}
		a\left[ \frac{1}{ m(R-R_o)+R_{co}}-\frac{s}{R} \right] & foam
\end{subnumcases}	
as ordinary differential \textit{average coarsening equations} for the evolution of the average radius $R(t)$ alone, decoupled from the critical radius $R_c(t)$.  Besides a materials-dependent rate factors, $a_w$ or $a$, the evolution of $R(t)$ depends on the two dimensionless parameters $m$ and $s$, and on initial conditions through $R_o$ and $R_{co}$. In effect, there are thus five distinct constants in the model.


\section{Solution of the Average Coarsening Equation for Dry Foam}

For later comparison with exact solution, it's helpful to consider a couple benchmarks.  First and most simply, if the sample is prepared in the scaling state, \textit{i.e.}~with $R_{co}=mR_o$, then the right-hand sides of Eqs.~(\ref{eq_dR}) are powers of $1/R$ and the solutions are powers of $t+\tau$ where $\tau$ is an integration constant.  Matching to the initial conditions, $R(t_o)=R_o$, the predicted scaling-state solutions are thus the familiar power laws
\begin{subnumcases}{\label{eq_Scaling} R(t)=}
	\label{eq_ScalingFroth}
		 \left[ \mathcal{D}(t-t_o) + {R_o}^3 \right]^{\frac{1}{3}} & froth \\
	\label{eq_ScalingFoam}
		\left[ \mathcal{D}(t-t_o) + {R_o}^2 \right]^{\frac{1}{2}}  & foam
\end{subnumcases}	
where the quantity $\mathcal{D}$ is given by
\begin{subnumcases}{\label{eq_D} \mathcal{D} = }
	\label{eq_Dfroth}
		  3\frac{a}{m}\left( 1-ms\right) & froth \\
	\label{eq_Dfoam}
		  2\frac{a}{m}\left( 1-ms\right) & foam
\end{subnumcases}	
and has units of a diffusion coefficient for dry foams.  For both cases, the corresponding critical radii are given by Eq.~(\ref{eq_RcVersusR}) as $R_c(t)=mR(t)$.  

If the sample is prepared close the scaling state, so that the transient difference of $R_c(t)$ from $mR(t)$ is small compared to $R(t)$, then we can find a perturbative solution where the leading difference from the scaling state is an additive constant $\delta$. In particular, if
\begin{subnumcases}{\label{eq_Pert} R(t) = }
	\label{eq_PertFroth}
		 \left[ \mathcal{D}(t-t_o) + {(R_o-\delta)}^3 \right]^{\frac{1}{3}} +\delta & froth  \\
	\label{eq_PertFoam}
		 \left[ \mathcal{D}(t-t_o) + {(R_o-\delta)}^2 \right]^{\frac{1}{2}} +\delta  & foam
\end{subnumcases}	
are inserted into the left and right hand sides of the average coarsening equations, and expanded in $\delta$, then equality holds to $\mathcal{O}(\delta^2)$ if the length scale $\delta$ is taken as
\begin{subnumcases}{\label{eq_delta} \delta = }
	\label{eq_deltaFroth}
		 \frac{R_o - R_{co}/m}{2(1-ms)} & froth \\
	\label{eq_deltaFoam}
		 \frac{R_o - R_{co}/m}{1-ms} & foam
\end{subnumcases}	
As a quick check, note that $\delta$ vanishes and Eqs.~(\ref{eq_ScalingFroth},\ref{eq_ScalingFoam}) are recovered if the sample is prepared in the scaling state. Note too that $\delta$ is negative if the initial preparation is too monodisperse, and is positive if it is too polydisperse.  Thus, $\delta$ emerges from the model as a signed length scale correlated to the width of the initial size distribution relative to that in the self-similar scaling state.


For a general initial sample preparation, exact solution of Eqs.~(\ref{eq_dR}) can be found for time versus radius by separating variables, integrating, and matching initial conditions.  For the dry foam case this gives
\begin{widetext}
\begin{eqnarray}
t-t_o  &=& \frac{(R-R_o)\left[m(1-ms)R-m(1+ms)R_o+2R_{co} \right] }{2a(1-ms)^2 }
+\frac{s(mR_o-R_{co})^2}{a(1-ms)^3}\ln\left[ \frac{(1-ms)R + s(mR_o-R_{co})}{R_o-sR_{co}} \right]
\label{eq_TvsR}  \\
  &=& \frac{R_oR_{co}(R-R_o)}{a(R_o-sR_{co})} + \frac{ (mR_o^2-s{R_{co}}^2)(R-R_o)^2}{2a(R_o-sR_{co})^2}
   + \frac{ s(mR_o-R_{co})^2(R-R_o)^3}{3a(R_o-sR_{co})^3}
 +\mathcal{O}\left( (R-R_o)^4 \right) \label{eq_TSmallVsR}
\end{eqnarray}
\end{widetext}
where the first equation is exact and the second is its expansion in $R$ around $R_o$.  As a first check on the exact result, it approaches the scaling state $t=R^2/[2a(1/m-s)]$ at very late times, where $t\gg t_o$ and $R\gg R_o$.  As a second check, expansion of the exact result to first order in $R_{co}$ around $mR_o$ is equivalent to simply dropping the log term; this gives the perturbative approximate solution of Eq.~(\ref{eq_PertFoam}).  The perturbative solution is also obtained by expanding the exact result in $1/R$ for any initial conditions; thus, Eq.~(\ref{eq_PertFoam}) can always be used to analyze data at late enough times even for far-from-scaling initial preparations.  For wet-froth limit of dilute bubbles, the exact solution of Eq.~(\ref{eq_dRfroth}) could be similarly computed and expanded.

\begin{figure}[t]
\includegraphics[width=3.2in]{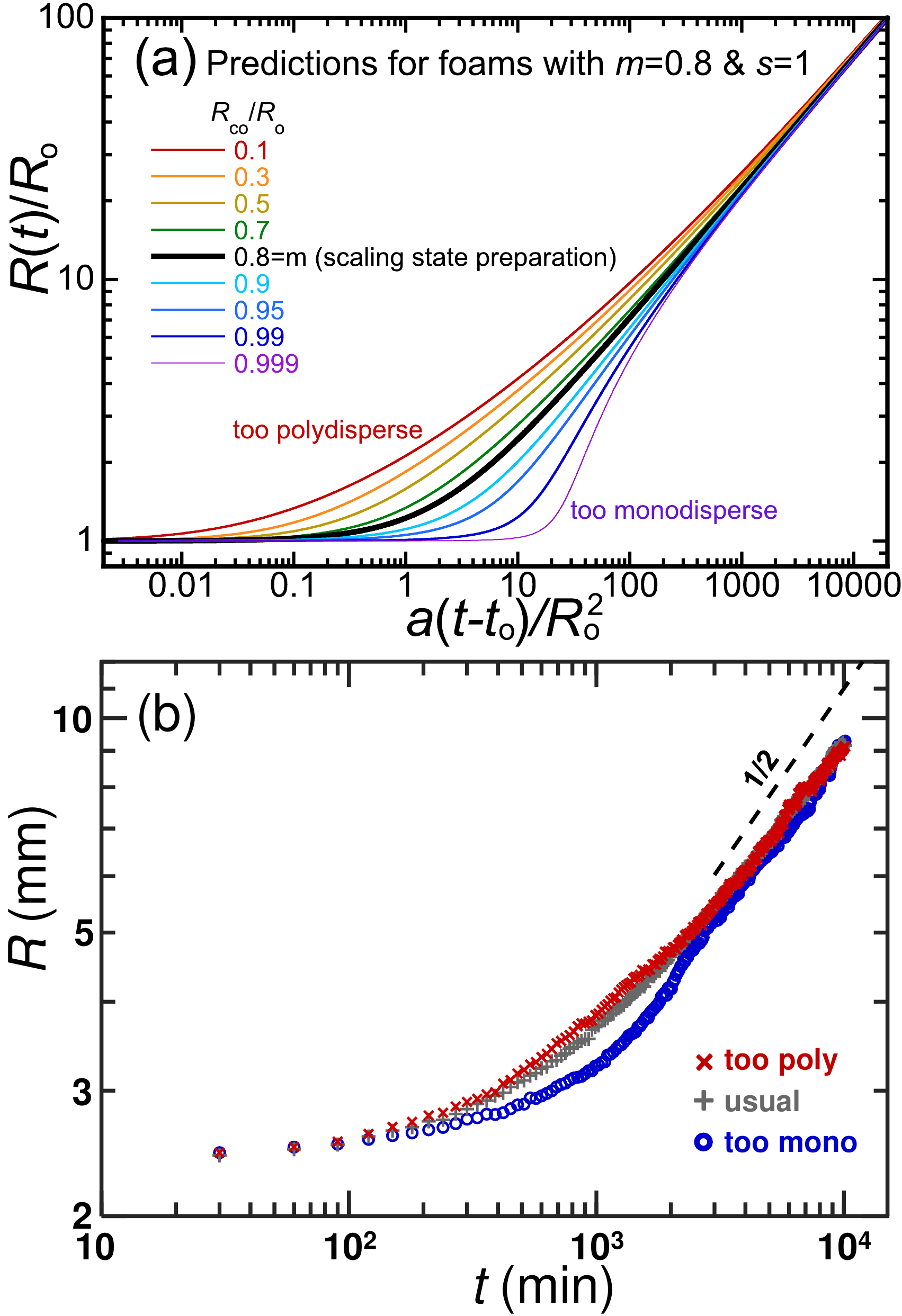}
\caption{Average bubble radius versus time (a) predicted by Eq.~(\ref{eq_TvsR}) for various parameters, as labeled, and (b) observed in experiment for three differently-prepared quasi-2d foam samples, as described in the text and pictured at time $t_o=30$~min. in Fig.~\ref{Fig_FoamImages}. The error bars are smaller than the symbol sizes.}
\label{Fig_RversusT}
\end{figure}

The predicted growth of the average radius for foams with different initial preparations is shown in Fig.~\ref{Fig_RversusT}a, where Eq.~(\ref{eq_TvsR}) is plotted for several $R_{co}$ values with arbitrary choices of $s=1$ and $m=0.8$, such that the scaling-state preparation is $R_{co}/R_o=0.8$. Note that Eq.~(\ref{eq_TvsR}) can be made dimensionless by multiplying each side by $a/R_o^2$, so that the left-hand side is $a(t-t_o)/R_o^2$ and so that on the right-hand side all factors of $R$ and $R_{co}$ are divided by $R_o$.  Thus we plot $R(t)/R_o$ versus $a(t-t_o)/R_o^2$ and need specify only the values of $R_{co}/R_o$, $m$, and $s$ -- but not $t_o$, $R_o$ or $a$.  For foams that are initially too polydisperse, with $R_{co}/R_o<m$, the average radius approaches the scaling state from above and the early-time growth is faster for larger polydispersity.  For foams that are initially too monodisperse, with $R_{co}/R_o>m$, the average radius approaches the scaling state from below.  And the initial growth is slower, approaching zero, for foam preparations that are progressively more monodisperse, $R_{co}\rightarrow R_o/s$ from below, since the pressure difference between neighboring bubbles begins small.  This whole phenomenology agrees with intuition, and is reminiscent of observations in Refs.~\cite{GlazierGrossStavans87, MarderPRA87, BeenakkerPRA88, GlazierGraner10, ZimnyakovCSA2019}. The trends also appear to coincide with average bubble growth for our own experiments, shown in Fig.~\ref{Fig_RversusT}b and described below.

As an aside, it should be pointed out that the average coarsening equations cannot account for the evolution of a spatially-varying initial size distribution, such as a single large bubble in an otherwise monodisperse but finite lattice, or a disordered sample where the average bubble size varies with position.  It also cannot account for evolution that does not reach a scaling state, such as a single large bubble in otherwise monodisperse but infinite lattice, or a perfectly ordered bidisperse lattice.

\section{Experiments}

Quasi-2d dry foams consisting of gas bubbles squashed between clear plates are ideal for testing our model for several reasons. First, they can be prepared with different initial bubble size and side-number distributions. Second, bubble areas can be extracted from images and circle-equivalent radii can be defined as $R_i=\sqrt{A_i/\pi}$. Third, the average critical radius can be found from the areas of the $n_i=6$ sided bubbles because, according to von~Neumann's law $dA_i/dt=K_o(n_i-6)$ \cite{Mullins56, VonNeumann}, they neither grow nor shrink.

\begin{figure*}[t]
\includegraphics[width=5in]{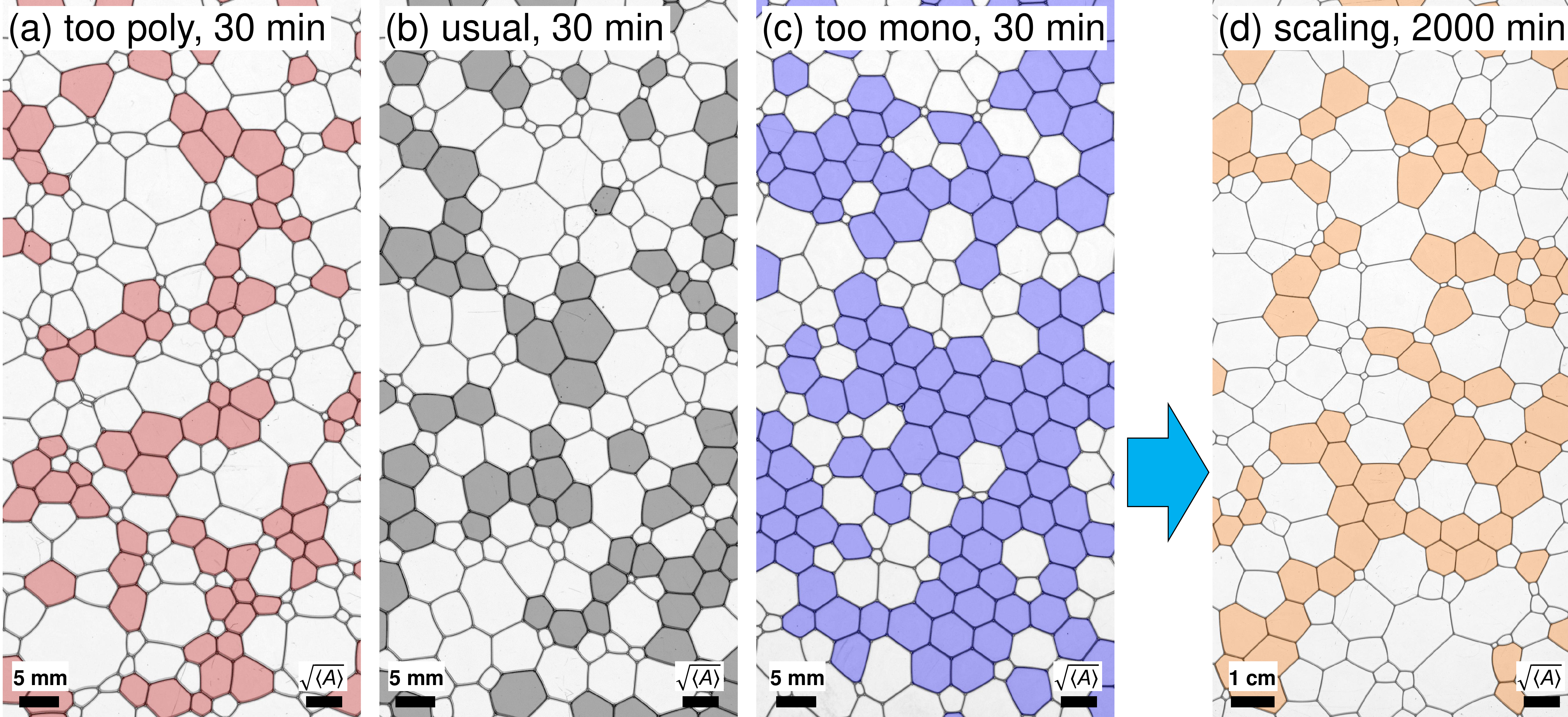}
\caption{(a-c) Images of foams with different initial preparations, corresponding to $t_o=30$~min.\ in Fig.~\ref{Fig_RversusT}, and (d) a foam in the self-similar scaling state much later. The critical six-sided bubbles are shaded; Those with more sides grow, while those with fewer sides shrink.  All images are $10 \times 20 \langle A\rangle$ in size, with the root mean bubble area $\sqrt{\langle A \rangle}$ used as scale bar.}
\label{Fig_FoamImages}
\end{figure*}

Our samples are made and measured as follows. The foaming solution is 80\% deionized water and 20\% Dawn Ultra Concentrated Dish Detergent. The sample cell consists of two 1.91~cm-thick acrylic plates separated by a spacing of 0.24~cm and sealed with two concentric o-rings, the inner of which has a 23~cm diameter, as in Refs.~\cite{RothPRE13, ChiecoPRE21a, ChiecoPRE21b, ChiecoFSM22}.  There is a trough surrounding the field of view, which is entirely filled with foaming solution. Then the cell is flushed with Nitrogen and sealed. For the ``usual" initial preparation, following prior works, the entire cell is vigorously shaken to produce bubbles that are homogeneously small compared to the gap between the plates.  For the ``too mono" preparation, Nitrogen is bubbled into the solution in the cell prior to sealing. This gives bubbles that are larger and more monodisperse than desired. So the sample is stirred a bit roughly with an enclosed magnetic disk in order to break up some of the hexagonal regions of bubbles. For the ``too poly" preparation, the cell is shaken but not very vigorously so that the foam is spatially nonuniform -- in some regions the bubbles are small compared to the gap while in others the bubbles are larger.  After a couple hours, these regions are gently stirred together with the enclosed magnetic mixer.  For the latter two cases, mixing introduces smaller than gap sized bubbles, which increases the polydispersity for both foams.

After preparation, the sample cell is placed 0.5~m above a Porta-Trace light box and 2~m below a Nikon~D850 camera with a Nikkor AF-S 300mm 1:2.8D lens and allowed to coarsen undisturbed. Images are then acquired every 30~minutes for at least 7 days. We only analyze images after the foams evolve into a quasi-2d state and the bubbles are large compared to the gap. Images are binarized, skeletonized and passed through a watershedding algorithm. From the watershed images we collect areas and number of sides for each bubble once the the sample has coarsened enough to become quasi-2d.  For the too-poly sample, this happens at about 7.5~hours after production; this defines $t_o$, which we take to be $t_o=30$~min since this is the time between images.  The other samples require different amounts of time to become quasi-2d. Therefore, to better compare, their time bases are shifted by an additive constant so that all three samples have approximately the same initial circle-equivalent average radius, $R_o = R(t_o) \approx 2.45$~mm at $t_o$. Zoomed-in images of some bubbles for all three samples at $t_o=30$~min are collected in Fig.~\ref{Fig_FoamImages}. There, the critical $n=6$ sided bubbles are highlighted in color. The ``too mono" sample clearly has a larger area fraction of critical bubbles, which are more similar in size.  The ``too poly" sample has broader distribution of sizes, than the ``usual" sample, but this is not as obvious.

\begin{figure}[t]
\includegraphics[width=3.2in]{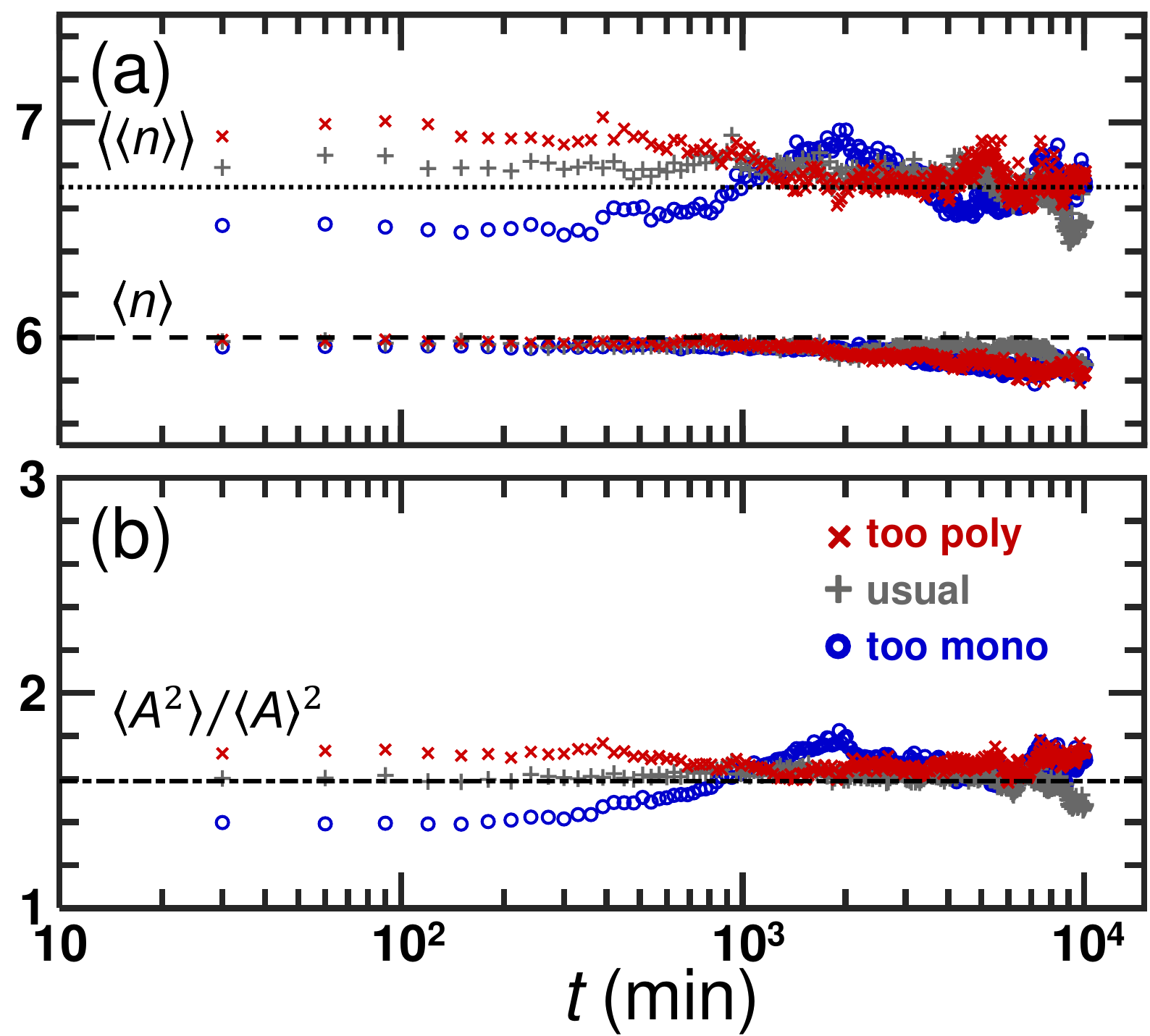}
\caption{Time dependence of (a) area-weighted average side number $\langle\langle n\rangle\rangle$ and average side number $\langle n\rangle$, plus (b) second moment divided by the average area squared. Symbols indicate data from foam samples with different initial conditions, as labeled. The black dashed lines show either $\langle n \rangle =6$ or the time average of ``usual" data for $\langle\langle n\rangle\rangle$ and $\langle A^2\rangle / \langle A\rangle^2$.}
\label{Fig_ScalingMeasures}
\end{figure}

\begin{figure}[t]
\includegraphics[width=3.2in]{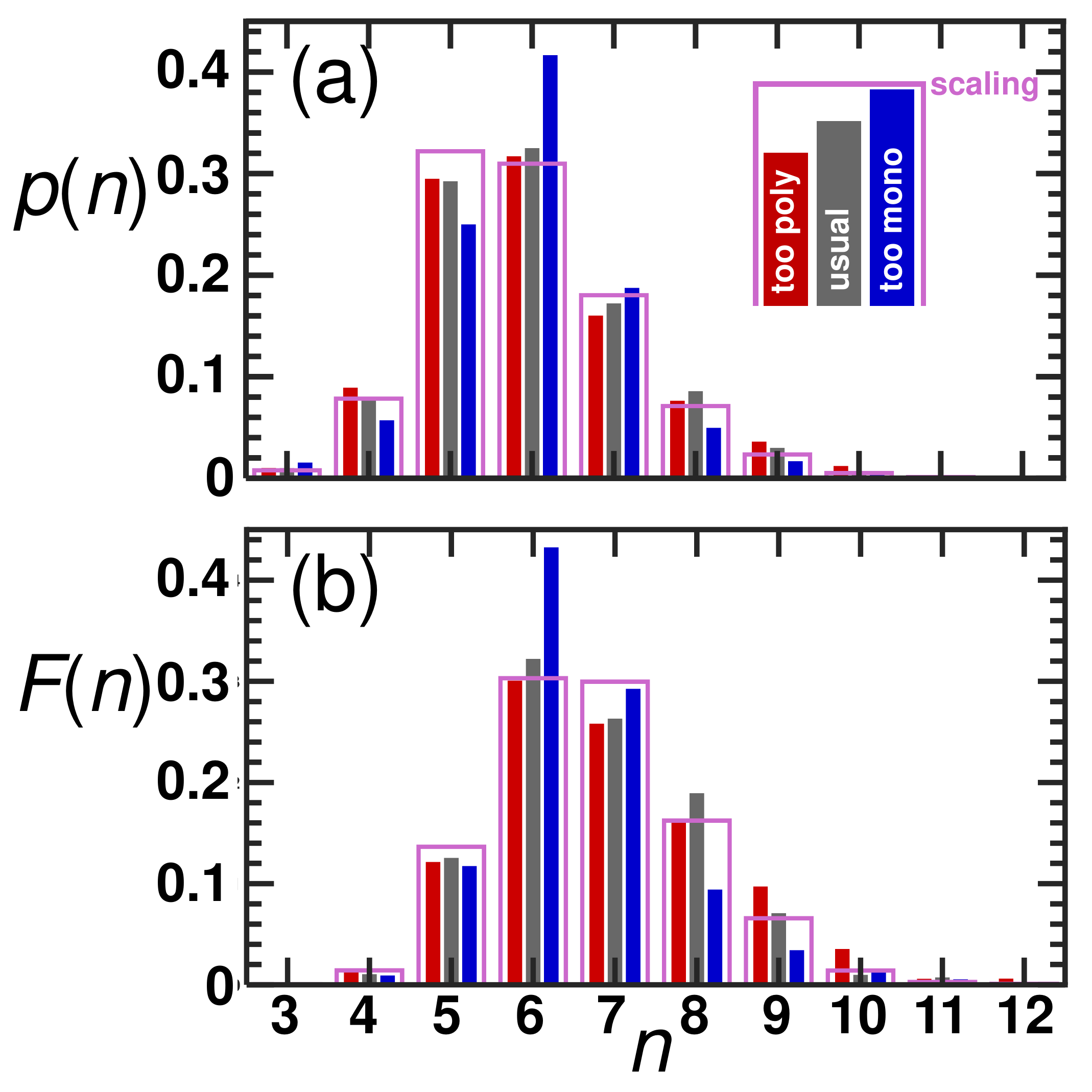}
\caption{(a) Side number distribution and (b) area-weighted side number distribution for the three foam preparations.  The narrow solid bars represent the three preparations at initial time $t_o$, while the wider outline represents their late time scaling-state average.}
\label{Fig_SideNumbers}
\end{figure}

Before fitting the $R(t)$ data in Fig.~\ref{Fig_RversusT}b to Eq.~(\ref{eq_TvsR}), we perform some important checks and auxiliary measurements. First, we verify that all three samples approach the same scaling state. Indeed, Fig.~\ref{Fig_RversusT}b shows that the average radii for the different preparations all converge at late times and grow thereafter consistent with $R\sim t^{1/2}$ scaling. Furthermore, Fig.~\ref{Fig_ScalingMeasures} shows the time evolution of three statistical measures of the bubble size and side-number distributions.  This includes the average side number $\langle n\rangle$, which equals six for an infinite sample by the Plateau and Euler laws \cite{WeaireRivier1984, WeaireHutzlerBook, Cantat2013, Langevin2020}; the area-weighted average side number $\langle\langle n\rangle\rangle$, whose difference from six controls the average coarsening rate in the scaling state \cite{Stavans90, ChiecoFSM22}; and the dimensionless second moment of the bubble area distribution $\langle A^2\rangle/\langle A\rangle^2$, which also affects the average coarsening rate. The latter two start at different values for all three samples, but converge together at late times -- thus demonstrating that a common scaling state has been reached. At the very latest times, however, all measures begin to deviate from a constant value.  This happens for $\langle n\rangle$, too, which falls below six as indicative of finite-size effects of no longer having a large number of bubbles in the sample. As a related check, we verify that the side number distributions are different for the three preparations at early times, as shown by the solid bars in Fig.~\ref{Fig_SideNumbers}.  At late times, consistent with an approach to a statistically self-similarity scaling state, the distributions become equal to within statistical uncertainty; these are averaged together and shown in Fig.~\ref{Fig_SideNumbers}.

\begin{figure}[t]
\includegraphics[width=3.2in]{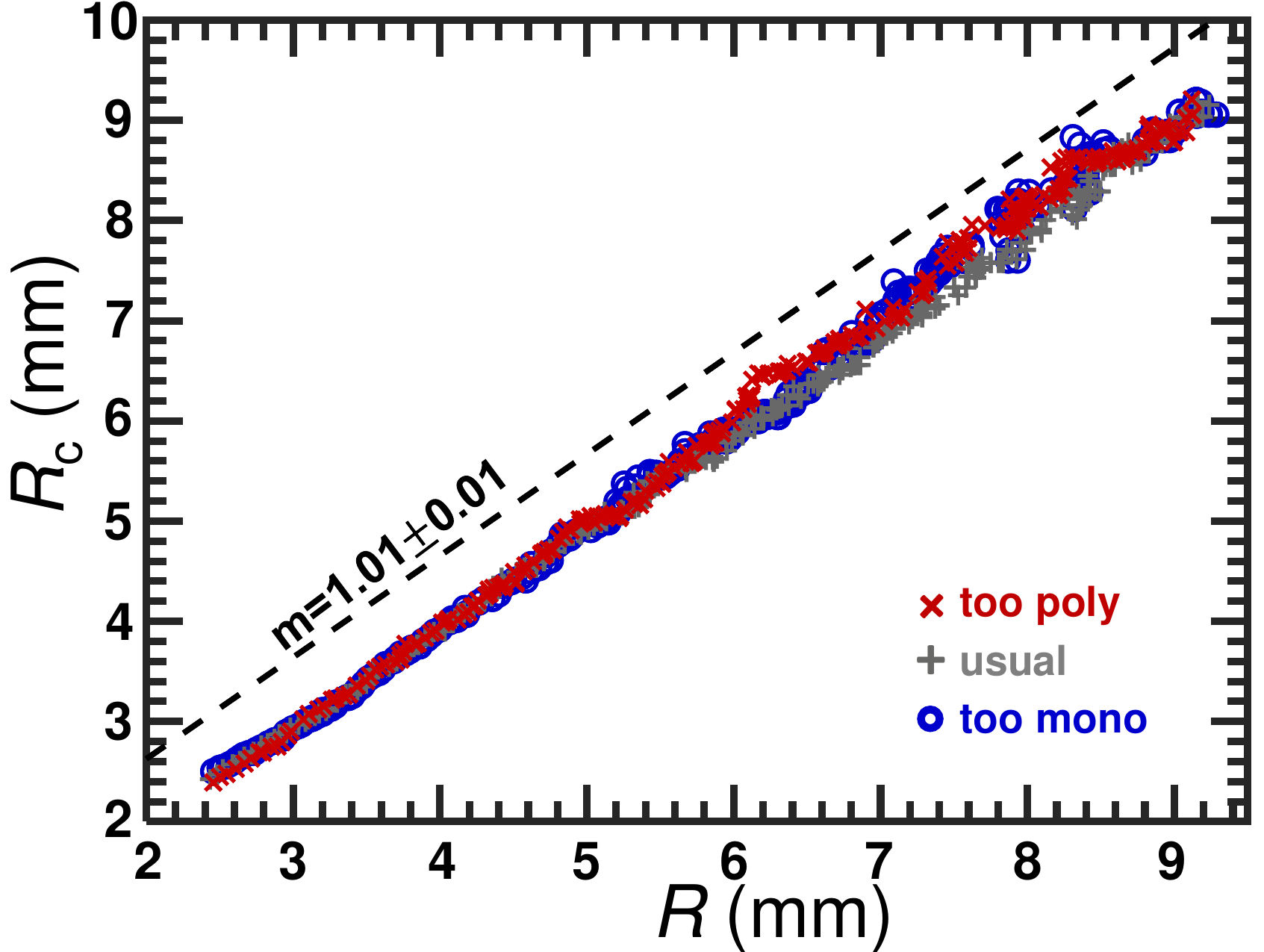}
\caption{The average radius of critical $n=6$ sided bubbles versus the average radius of all bubbles in the coarsening foams, plotted parametrically during evolution from $t_o=30$~min to the end of the experiment at $t\approx10^4$~min. The different symbols indicate different initial preparations. The three data sets are fit to a line $R_c=mR+b$, giving $m=1.01\pm0.01$ as a simultaneous fit parameter and different values of $b$ for each data set.}
\label{Fig_RCvsR}
\end{figure}

The last and perhaps most crucial check is to investigate the validity of a key assumption of the model.  Namely, Eqs.~(\ref{eq_mdef}-\ref{eq_RcVersusR}) presume that as the critical radius $R_c(t)$ grows, it remains linear in the average radius $R(t)$. Thus, in Fig.~\ref{Fig_RCvsR}, we plot the average circle-equivalent radius of the six-sided bubbles parametrically versus the average circle-equivalent radius of all bubbles. There indeed we find a linear relationship, with a common slope, $m$, as assumed. Simultaneous fits to the three different sample preparations gives $m=1.01\pm0.01$, as illustrated by an offset dashed line with this slope.  The uncertainty in $m$ spans acceptable values for fits over the whole range as well as restrictions to $R<7$~mm and $R>5$~mm.

\begin{figure}[t]
\includegraphics[width=3.2in]{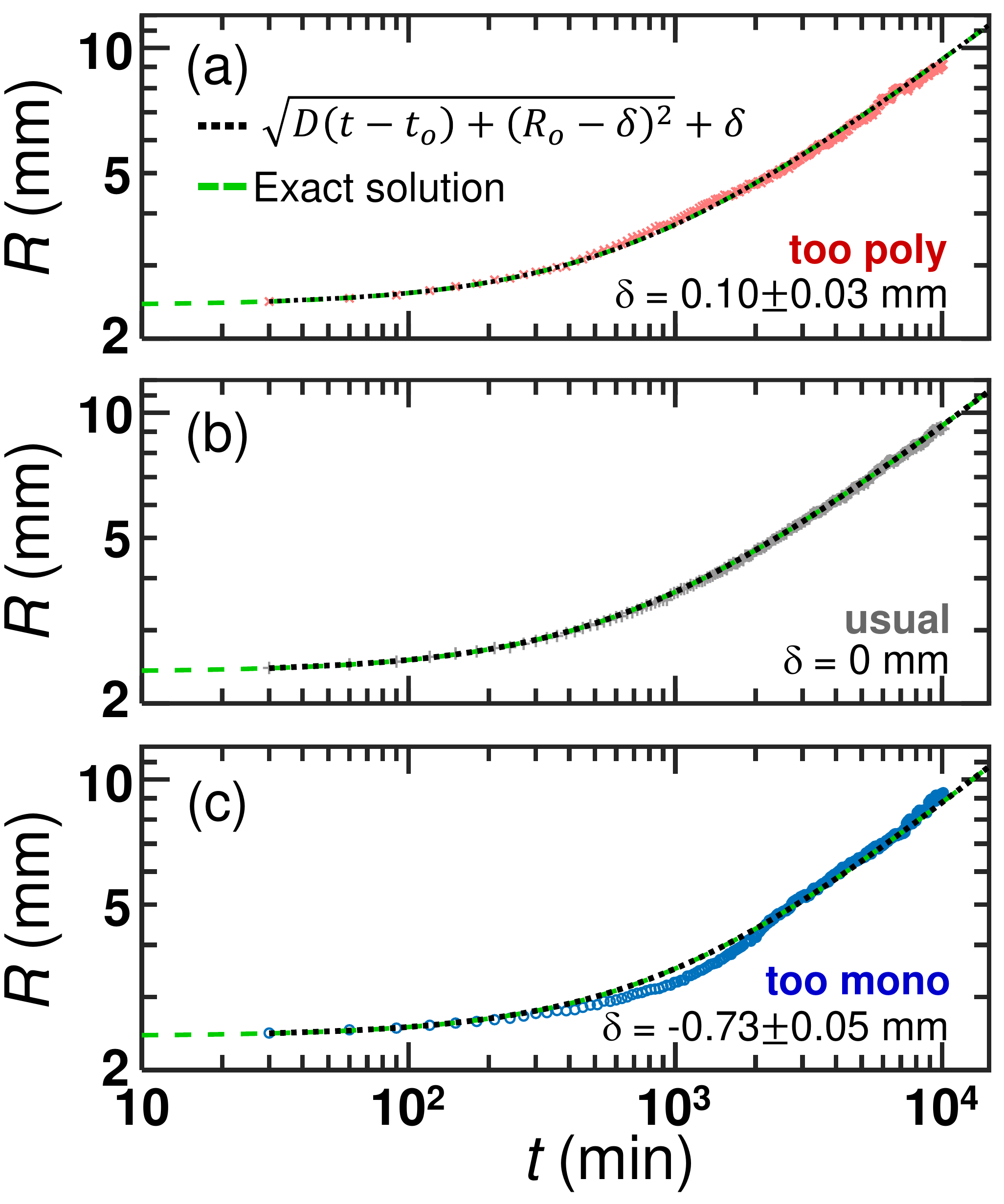}
\caption{Average radius versus time for foams prepared with different initial conditions, as labeled. The symbols represent experimental data. The dashed curves show fits to Eq.~(\ref{eq_TvsR}) where $s$ and $a$ are used as simultaneous fit parameters.  The values of $R_o$ and $R_{co}$ are fixed separately for each data set, while $m=1.01$ is fixed from the analysis shown in Fig.~\ref{Fig_RCvsR}.  The dotted curves result from dropping the log term in the fitting function, giving $\delta$ values from Eq.~(\ref{eq_deltaFoam}) as shown in the plots. }
\label{Fig_RvsTfits}
\end{figure}

Now we are well-positioned to compare the prediction of Eq.~(\ref{eq_TvsR}) with data for the growth of the average bubble radii, $R(t)$, for the three preparations (Fig.~\ref{Fig_RvsTfits}). Since the initial average $R_o$ and critical $R_{co}$ radii are known at time $t_o=30$~min,  and since $m=1.01\pm0.01$ is fixed from Fig.~\ref{Fig_RCvsR}, the only unknown parameters of the model are $s$ and $a$. To find these, we simultaneously fit to $R(t)$ data for all three sample preparations.  This gives $s=0.954\pm 0.009$ and $a=0.123\pm 0.006$, which combine with $m$ to give $\mathcal{D}=0.009\pm 0.003$~mm$^2$/min from Eq.~(\ref{eq_D}b) as well as $\delta$ values from Eq.~(\ref{eq_deltaFoam}) shown in the figure. The uncertainties in these quantities reflect both the error from the fit as well as the span of values coming from the range of $m=1.01\pm0.01$. The resulting fits, shown as dashed curves in Fig.~\ref{Fig_RvsTfits}, agree well with the data. However, close inspection reveals a small systematic discrepancy.  This is largest for the ``too mono" sample, which crosses over to the scaling form faster than the model predicts.

Note that dropping the log term of Eq.~(\ref{eq_TvsR}), \textit{i.e.} plotting the perturbative solution Eq.~(\ref{eq_PertFoam}) using $\mathcal{D}$ and $\delta$ values given respectively by Eqs.~(\ref{eq_Dfoam},\ref{eq_deltaFoam}) with the above parameters, gives agreement that is essentially unchanged (dotted curves in Fig.~\ref{Fig_RvsTfits}). An alternative analysis is thus to do simultaneous fits to the perturbative solution; this gives $\mathcal{D}=0.0084 \pm 0.0004$~mm$^2$/min, which is consistent with the above results but has smaller uncertainty than from propagating the uncertainties in $m$ and $s$. The fitting values for $\delta$ are $0.09\pm 0.03$~mm and $-0.79\pm 0.06$~mm for the too poly and too mono cases, respectively, similarly consistent with the results computed from fits to the full form. Thus, the perturbative solutions is a good approximation to the full solution, consistent with the expected sign of $\delta$ and with $|\delta|$ being fairly small compared to $R_o$.

As an aside, note that average coarsening rate is controlled by $\mathcal{D} \propto (1-ms)$, which is small because $m$ and $s$ are both close to one. This is perhaps analogous to the relation $d\langle A\rangle/dt \propto [\langle\langle n\rangle\rangle -6]$ noted earlier for ideal dry 2d foams, which is small because the area-weighted average side number is close to the average side number.

\section{Prior Data Revisited}

\begin{figure}[t]
\includegraphics[width=3.2in]{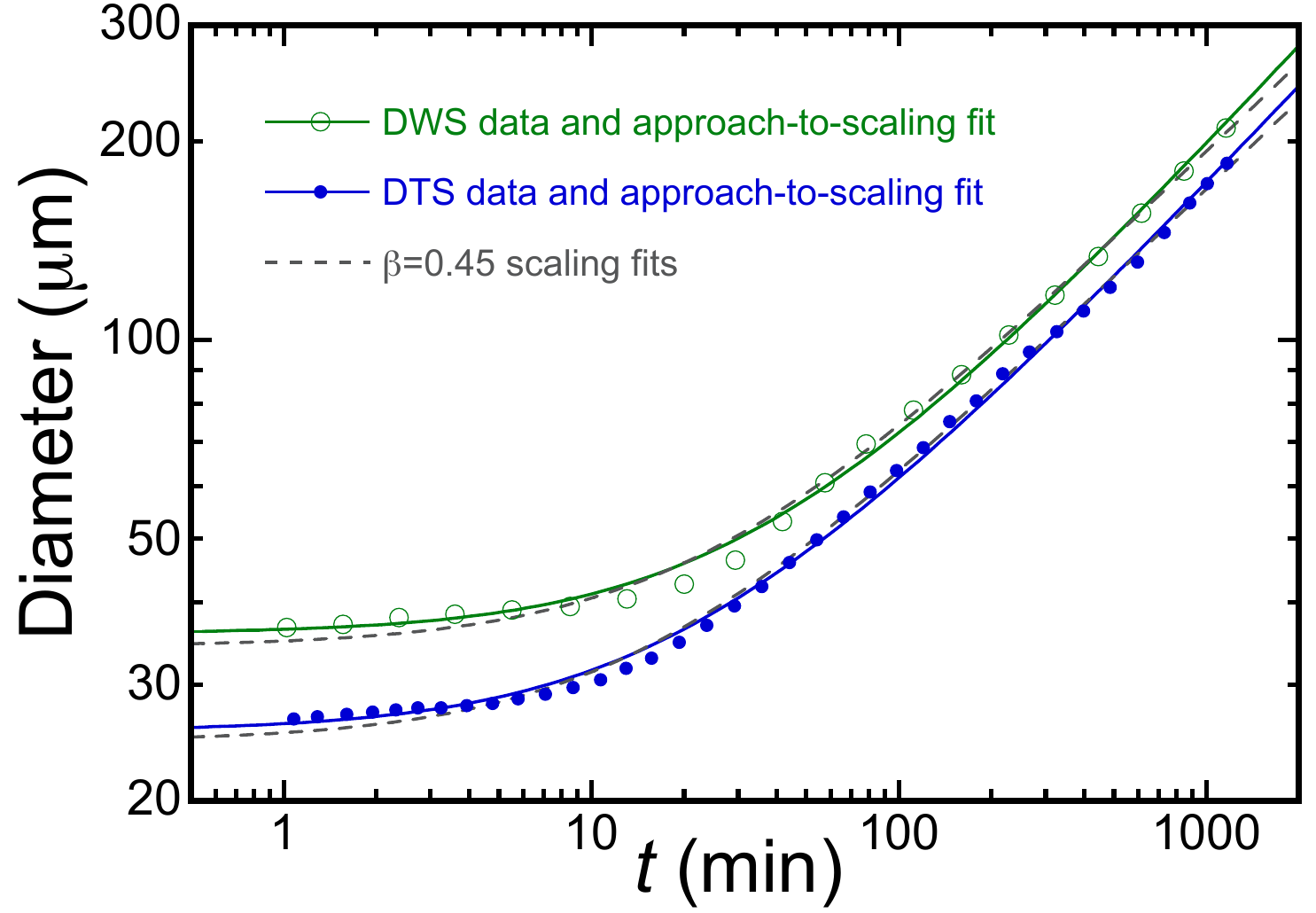}
\caption{Average measures of bubble diameter in a shaving cream, extracted from Fig.~3 of Ref.~\cite{DurianWeitzPine91b} and based on diffusing-wave spectroscopy (DWS) and diffuse transmission spectroscopy (DTS) probes of bulk behavior. The dashed curves are fits to Eq.~(\ref{eq_scalingbeta}) with fixed exponent $\beta=0.45$.  The solid curves are fits to Eq.~(\ref{eq_approach}), which gives exponent values indistinguishable from $\beta=1/2$.  This is consistent with the dry foam limit and a non-scaling state initial preparation that is too polydisperse.}
\label{Fig_PRA}
\end{figure}

In this last section we demonstrate use of our model for analyzing systems with transients or apparent late-time power-law growth falling between the known extremes of $\beta=1/2$ and $\beta=1/3$.  For foams, Refs.~\cite{Isert2013, CRAS2023, Durian2023, esacoarsening} concern the systematic variation of $\beta$ with liquid content.  In such cases it is common practice to deduce an empirical coarsening exponent, $\beta$, for example by fitting average radii or diameter data to a function of form
\begin{equation}
    R(t) = \left[ \mathcal{D}(t-t_o)+R_o^{1/\beta} \right]^\beta
\label{eq_scalingbeta}
\end{equation}
This assumes -- without justification -- that the system was prepared in a scaling regime, \textit{i.e.} that there are no confounding effects due to the decay of unknown transients. For example, Fig.~3 of Ref.~\cite{DurianWeitzPine91b} shows two measures of an average diameter for a shaving cream with 8\% liquid that appears to approach $\beta=0.45\pm0.05$ at late times.  It's unclear whether this is consistent with $\beta=1/2$ or if the shaving cream is outside the dry limit and has a nontrivial effective exponent that is truly less than one~half.  We now have a tool to study this.  First, diameter data were digitized from Ref.~\cite{DurianWeitzPine91b} and fit to Eq.~(\ref{eq_scalingbeta}) with $\beta=0.45$ fixed according to the reported late-time power-law behavior.  The fits, shown by dashed curves in Fig.~\ref{Fig_PRA} are good at late times but not very satisfactory at early times.  The same data are now fit to a form similar to Eqs.~(\ref{eq_Pert}) but with a variable exponent:
\begin{equation}
    R(t) = \left[ \mathcal{D}(t-t_o)+(R_o-\delta)^{1/\beta} \right]^\beta + \delta
\label{eq_approach}
\end{equation}
We take $t_o=1$~min and adjust the other three parameters to match the data. The fits, shown by solid curves in Fig.~\ref{Fig_PRA}, are more satisfactory over the whole range of data.  These give positive $\delta$ values of $7.1\pm0.6$~$\mu$m for both samples, which is almost four times smaller than the diameter at time $t_o$ and is consistent with an initial preparation that is slightly more polydisperse than the scaling state.  Furthermore, the resulting effective exponents are $0.52\pm0.08$ for the DTW measure and $0.51\pm0.05$ for the DTS measure.  These $\beta$ values are consistent with the dry limit expectation.  In fact, when the fits to Eq.~(\ref{eq_approach}) are repeated with fixed $\beta=1/2$, the resulting curves are indistinguishable from those shown. For all these reasons, we conclude that the Ref.~\cite{DurianWeitzPine91b} shaving cream is consistent with the usual dry limit power-law but having an initial polydispersity that is initially slightly greater than that in the long time scaling state.

\begin{figure}[t]
\includegraphics[width=3.2in]{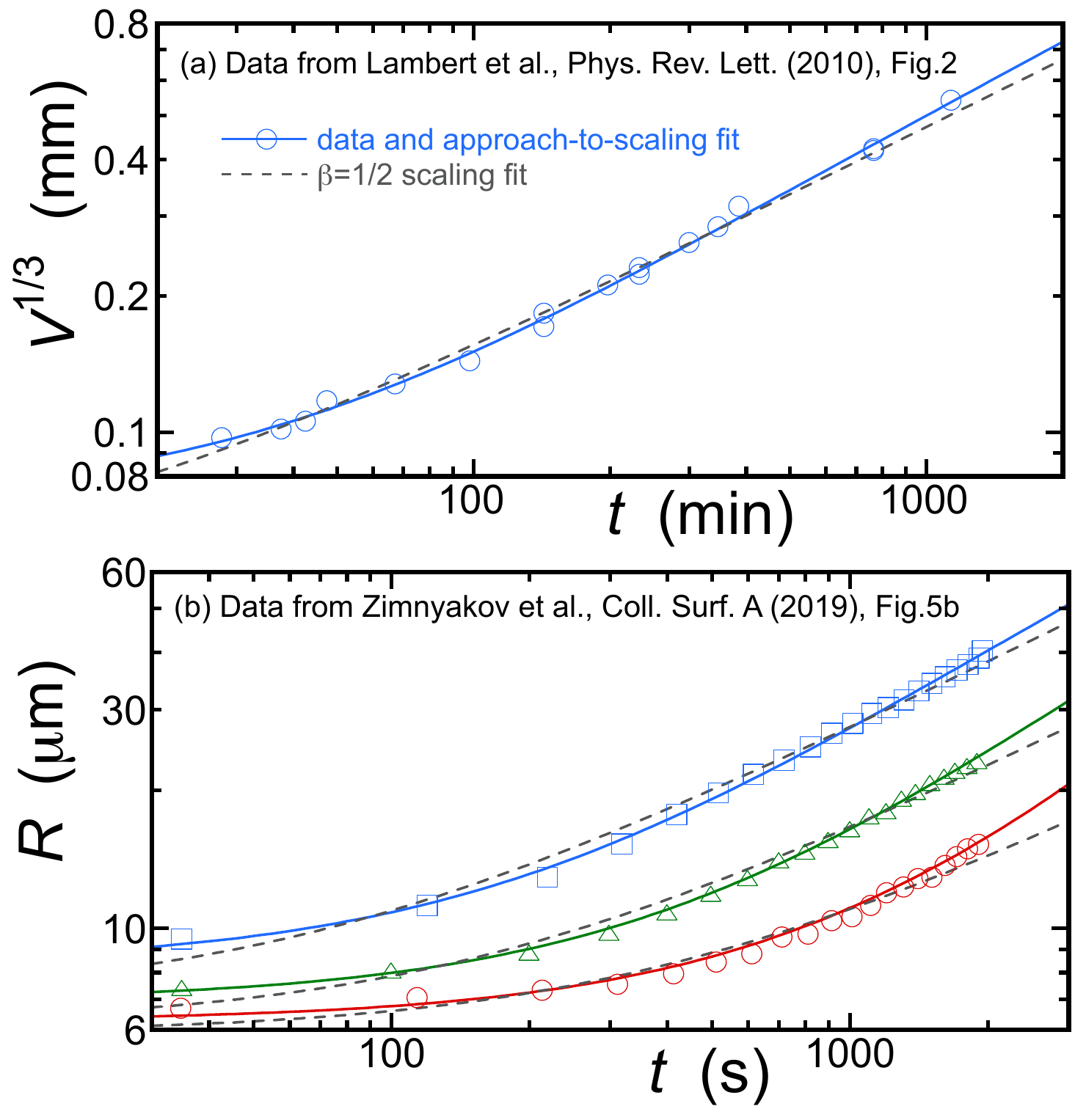}
\caption{Bubble size data digitized from (a) Ref.~\cite{GlazierGraner10} and (b) Ref.~\cite{ZimnyakovCSA2019}.  The solid curves are fits to Eq.~(\ref{eq_PertFoam}), which accounts for transients, and dashed curves are fits to the scaling form Eq.~(\ref{eq_ScalingFoam}), which assumed that the foams are in the scaling state.}
\label{Fig_PriorData}
\end{figure}

A similar demonstration is made in Fig.~\ref{Fig_PriorData}, which shows (a) x-ray tomography data for a 3d dry foam \cite{GlazierGraner10} and (b) surface image data for a shaving cream at three different temperatures \cite{ZimnyakovCSA2019}.  These two papers specifically conclude that the systems reach a long time scaling regime with $\beta=1/2$. So we fit to the Eq.~(\ref{eq_PertFoam}) approximate solution of our average coarsening equation to test how well it captures the transients. For these fits we fix $t_o$ to the minimum of the $x$-axis and we fit for $\mathcal{D}$, $R_o$, $\delta$.  As seen in Fig.~\ref{Fig_PriorData}, all four fits to our model are outstanding (solid curves).  By contrast fits to Eq.~(\ref{eq_ScalingFoam}) for $\mathcal{D}$ and $R_o$, assuming a self-similar scaling state preparation, all show systematic deviation from the data (dashed curves).  Though our model has one more fitting parameter, and hence can be expected to fit better, this analysis lends support to both the authors' conclusions as well as the validity of our model.

\section{Conclusion}

In summary, by approximating the full domain size distribution in terms of the average and critical radii, we constructed a model for average growth that is both intuitive and solvable. For the dry foam case of nearly space-filling domains, we showed that it is supported by specially-designed experimental tests as well as by comparisons with prior data for the decay of transients from the initial state toward a long time scaling state. For the wet froth case of dilute bubbles, we showed that the approximate perturbative solution has the same form, \textit{i.e.} Eq.~(\ref{eq_approach}) but with $\beta=1/3$ rather than $\beta=1/2$. This equation features an interesting length scale, $\delta$, which emerges as a measure of the width of the initial domain size distribution relative to that in the scaling state.  Both $\delta$ and $\beta$ could now be treated as adjustable parameters for empirical data analysis, as demonstrated in Fig~\ref{Fig_PRA}.  We hope these contributions will be helpful for analyzing coarsening behavior in disparate systems, by providing a simple means to account for the possibly-confounding effects of unknown transients due to non-scaling state preparations. It would be interesting if our model could be derived, or if bounds on its accuracy could be set, from the partial integrodifferential equations for the wet \cite{Lifshitz1961, Wagner1961} and dry \cite{Lemlich1978, Wagner1961} limits.

\begin{acknowledgments}
This work was supported by NASA grant 80NSSC19K0599.
\end{acknowledgments}

\bibliography{FoamRefs}

\end{document}